# MCA BASED PERFORMANCE EVALUATION OF PROJECT SELECTION

Tuli Bakshi[1] and Bijan Sarkar[2]

[1]Department of Computer Application, Calcutta Institute of Technology, Howrah, India
`tuli.bakshi@gmail.com`
[2] Department of Production Engineering, Jadavpur University, Kolkata, India
`bijon_sarkar@email.com`

## ABSTRACT

*Multi-criteria decision support systems are used in various fields of human activities. In every alternative multi-criteria decision making problem can be represented by a set of properties or constraints. The properties can be qualitative & quantitative. For measurement of these properties, there are different unit, as well as there are different optimization techniques. Depending upon the desired goal, the normalization aims for obtaining reference scales of values of these properties. This paper deals with a new additive ratio assessment method. In order to make the appropriate decision and to make a proper comparison among the available alternatives Analytic Hierarchy Process (AHP) and ARAS have been used. The uses of AHP is for analysis the structure of the project selection problem and to assign the weights of the properties and the ARAS method is used to obtain the final ranking and select the best one among the projects. To illustrate the above mention methods survey data on the expansion of optical fibre for a telecommunication sector is used. The decision maker can also used different weight combination in the decision making process according to the demand of the system.*

## 1. INTRODUCTION

Being a temporary attempt, a project needs to create a unique product, service or result. Temporary signifies that a particular project has a definite dead line, reaching the dead line the project objectives has been gained or it becomes clear that the project objective will not be made or the necessity of the project no longer exists. In real world there can be multiple alternative projects. A decision maker (DM) has to choose one alternative which must be the best option. Therefore it is a very difficult task [1]. Selection and evaluation of a project involves decisions those are critical to profitability, growth and survival of organization in the competitive world. This type of decision involves multiple factors such as identification, considerations and analysis of viability. According to Hwang and Yoon [2] Multi-criteria decision making (MCDM) is applied to preferable decisions among available classified alternatives by multiple attributes. So MCDM is one of the most widely used decision methodology in project selection problems. The MCDM is a method that follows the analysis of several criteria, simultaneously. In this method economic, environmental, social and technological factors are considered for the selection of the project and for making the choice sustainable [3-5]. Several framework have been proposed for solving MCDM problems, namely Analytical Hierarchy Process[AHP] [6,7,8],Analytical Network Process[ANP] [9],which deals with decisions in absence of knowledge of the independence of higher level elements from lower level elements and about the independence of the elements within a level. Other framework available are data envelopment analysis ( DEA),Technique for order performance by similarity to ideal solution (TOPSIS) [10-11],VIKOR, COPRAS [12], with grey number,[13-15],Simple Additive weighting ( SAW) etc [16], LINMAP [17].With these techniques alternative ratings are measured, weight of the criteria are expressed in precise numbers [18]..The projects' life cycle assessment is to be determined and the impact of all actors is to be measured. There are some mandatory axioms that the criteria describing feasible alternatives are dimensions which are important to determine the performance.

## 2. TAXONOMY OF MCDM FOR PROBLEM SOLUTION

Evaluating a finite set of alternatives for finding the best one and to rank them from best towards, a decision maker, has to cluster them into predefine homogeneous classes. Pareto in 1986 [19] was the first to apply multi criteria optimization and determination of priority and utility function on problem set. Under pre-referential and utility independence assumption Keeny and Raiffa [20] offered the theorem for determining multiple criteria utility function. For solving problems with conflicting goals of global importance, Satty [21] presented decision making models with incomplete information.

In MCDM approach it is necessary to define the problem first and there after to identify realistic alternatives. It is very important to determine the actors involve in decision-making, evaluation criteria selection and evaluate all the alternatives according to the set of criteria. Guiton and Martel [22] gave an approach to select the appropriate MCDM method to a specific decision making situation.

Broadly MCDM methods are classified into two types- quantitative measurement and qualitative measurement. The method based on multi-criteria utility theory of first kind are TOPSIS (Technique for order preference by similarity to ideal solution) [23], SWA (Simple Additive Weighting), [24], LINMAP (Linear Programming Techniques for Multidimensional Analysis of Preference) [25], ARAS [26].

The second type is qualitative measurement. These include two widely known group of methods AHP [27-32] and Fuzzy set theory method [33].

## 3. METHODS: ADDITIVE RATIO ASSESSMENT (ARAS)

### 3.1. Step 1: - Establishment of Decision Making Matrix (DMM)

The first stage of ARAS method is decision making matrix (DMM) formation. In case of MCDM problem, the problem can be solved by representing the following DMM of preferences for m feasible alternatives (rows) and n sign full criteria (Columns) as:

$$
X = \begin{pmatrix}
x_{01} & \cdots x_{0j} & \cdots x_{0n} \\
x_{11} & \cdots x_{1j} & \cdots x_{1n} \\
\vdots & \cdots \vdots & \cdots \vdots \\
x_{i1} & \cdots x_{ij} & \cdots x_{in} \\
\vdots & \cdots \vdots & \cdots \vdots \\
x_{m1} & \cdots x_{mj} & \cdots x_{mn}
\end{pmatrix}
$$

Where i = No. of alternatives = 0, 2 ….m. and j = No. of criteria = 1, 2 …n. $x_{ij}$ = Score / performance value for i$^{th}$ alternative of j$^{th}$ criterion. And $x_{0j}$ = optimal value of the $j^{th}$ criterion, if optimal value of $j^{th}$ criterion is unknown, then $x_{0j}$ will be $(x_{ij})_{max}$ if the criterion is preferable.

$x_{0j}$ will be $(x_{ij})_{min}$ if the criterion is non-preferable. The performance values $x_{ij}$ and the criteria weights $w_j$ are viewed in the entries of a DMM. The weights of criteria are determined by the experts in AHP methods where $w_j$ = Weight / importance of $j^{th}$ criterion.

$$\sum_{j=1}^{n} wj = 1$$

## 3.2. Normalization of DMM

In the second stage, the initial values of all the criteria of the decision making matrix are normalized as:

$$\overline{X} = \begin{bmatrix} \overline{x_{01}} & \cdots \overline{x_{0j}} & \cdots \overline{x_{0n}} \\ \vdots & \ddots \vdots & \ddots \vdots \\ \overline{x_{i1}} & \cdots \overline{x_{ij}} & \cdots \overline{x_{in}} \\ \vdots & \ddots \vdots & \ddots \vdots \\ \overline{x_{m1}} & \cdots \overline{x_{mj}} & \cdots \overline{x_{mn}} \end{bmatrix}$$

Where $\overline{x_{ij}} = \dfrac{x_{ij}}{\sum_{i=0}^{m} x_{ij}}$, for benefit criteria.

The criteria whose preferable values are minima are normalized by applying two stage procedures as follows:

$$x_{ij} = \frac{1}{x_{ij}^{*}}$$

$$\overline{x_{ij}} = \frac{x_{ij}}{\sum_{i=0}^{m} x_{ij}}$$

## 3.3. Calculation of Criteria

Calculation of the importance of criteria by AHP / Logic Method / Modified Logic Method.

## 3.4. Calculation of Weighted Normalised Matrix

$$X = \begin{bmatrix} x_{01} & \cdots x_{0j} & \cdots x_{0n} \\ \vdots & \ddots \vdots & \ddots \vdots \\ x_{i1} & \cdots x_{ij} & \cdots x_{in} \\ \vdots & \ddots \vdots & \ddots \vdots \\ x_{m1} & \cdots x_{mj} & \cdots x_{mn} \end{bmatrix}, \text{ where } x_{ij} = \overline{x_{ij}} * w_j$$

### 3.5. Optimal Values

The optimal value is determined as follows:

$$S_i = \sum_{j=1}^{n} x_{ij}$$

where $S_i$ = value of the optimality function of $i^{th}$ alternative

### 3.6. Final Result

$K_i = S_i / S_0$, where $K_i$ = degree of utility for $i^{th}$ alternative and $S_0$ = the best or optimal one.

The largest value of $K_{i \, is}$ the best and the smallest one is the worst. Also the optimality function $S_i$ has a direct and proportional relationship with the values of $x_{ij}$ and weights $w_j$ and their relative influence on the final result.

## 4. PROPOSED MODEL

The proposed model for the project selection problem, composed of AHP and ARAS methods [34-35], consists of three basic stages: identification of properties, weight assigning and evaluation of alternatives and determine final rank. Based on proposed methodology, the present researcher selects some criteria like:

### 4.1. Net Present Value

The Net Present Value (NPV) is defined as the sum of the present values (PVs) of the individual cash flows. Actually NPV is an indicator of how much value a project adds to the organization. So it is treated as the benefit criteria of the project. In financial theory, if there is a choice between two mutually exclusive alternatives, the one yielding the highest NPV should be selected. So if the value of NPV is positive, the project may be accepted.

### 4.1. Rate of Return

Rate of return (ROR) is the ratio of money gained or lost on a project relative to the amount of money invested. ROR is usually expressed as a percentage. So ROR is also the benefit criteria for any project selection.

### 4.1. Payback Period

Payback period is the period of time required for the return on an investment or project. Payback period has no explicit criteria for decision making. Any project yielding the quickest Payback Period should be selected.

### 4.1. Project Risk

There may be some external circumstances or event that cannot occur for the project to be successful. The external events are called project risks. If such type event is likely to happen, then it would be a risk. The aim of project selection is to minimize the risk criteria.

After identifying these criteria, their weights are found by AHP method. Five homogeneous experts help us to specify the weight.

## 5. SCHEMATIC DIAGRAM OF PROPOSED MODEL

The schematic diagram of the proposed model is given below:

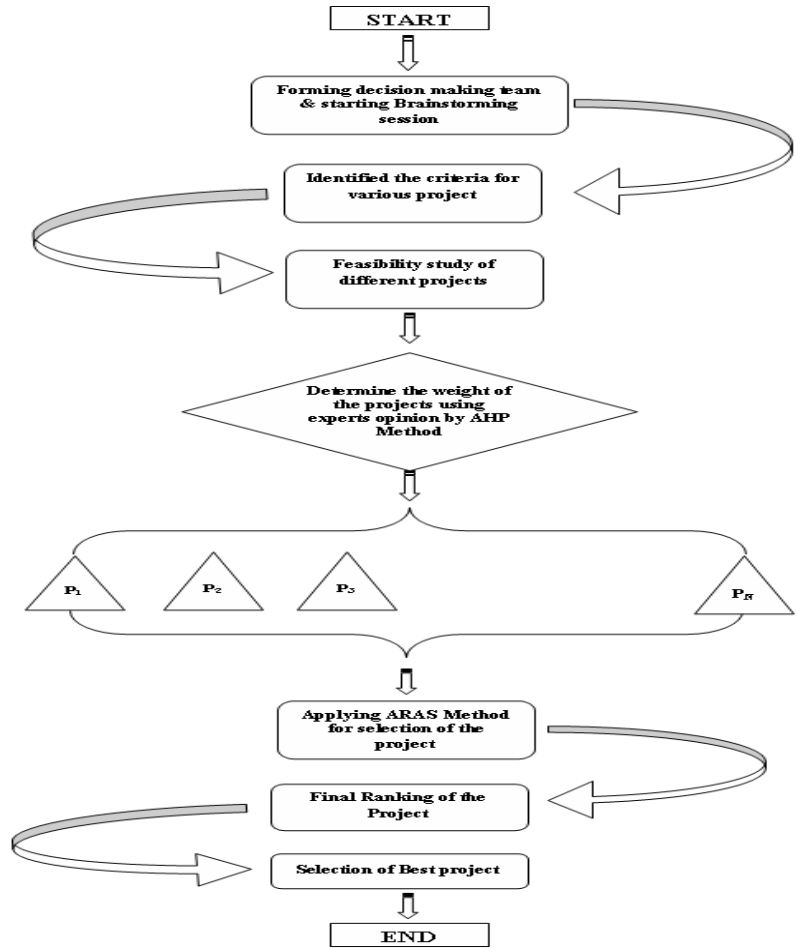

## 6. CASE STUDY OF PROPOSED MODEL

The survey data of the expansion of optical fibre for Telecommunication sector in one part of IRAN [36] is reused.

Table1. Problem Description Table for ARAS Method

| Serial No. | Set of criteria for evaluation | Variable | Optimal | Unit of Measurement | Weight |
|---|---|---|---|---|---|
| 1 | Net Present Value (NPV) | $X_1$ | MAX | Rs. (Rupees) | 0.29 |
| 2 | Rate of Return (ROR) | $X_2$ | MAX | Rs. (Rupees) | 0.34 |
| 3 | Payback Period (PB) | $X_3$ | MIN | Days (Month) | 0.22 |
| 4 | Project Risk (PR) | $X_4$ | MIN | − | 0.15 |

Table 2. Decision Matrix

| | NPV (+) | ROR (+) | PB (-) | PR (-) |
|---|---|---|---|---|
| Project 1 | 10 | 3 | 6 | 7 |
| Project 2 | 13 | 5 | 7 | 9 |
| Project 3 | 9 | 1 | 8 | 1 |
| Project 4 | 11 | 3 | 8 | 7 |
| Project 5 | 12 | 5 | 10 | 5 |

Table 3.  Normalized DMM

|  | NPV | ROR | PB | PR |
|---|---|---|---|---|
| Project 1 | 0.18 | 0.18 | 0.15 | 0.24 |
| Project 2 | 0.24 | 0.29 | 0.18 | 0.31 |
| Project 3 | 0.16 | 0.06 | 0.20 | 0.03 |
| Project 4 | 0.20 | 0.18 | 0.20 | 0.24 |
| Project 5 | 0.22 | 0.29 | 0.26 | 0.17 |

Table 4.

| Alternatives | Criteria | | | |
|---|---|---|---|---|
|  | $X_1$ | $X_2$ | $X_3$ | $X_4$ |
| Optimization Direction | MAX | MAX | MIN | MIN |
| Weight of criterion | **0.29** | **0.34** | **0.22** | **0.15** |
| $A_0$ | 1.00 | 1.00 | 0.15 | 0.03 |
| $A_1$ | 0.18 | 0.18 | 0.15 | 0.24 |
| $A_2$ | 0.24 | 0.29 | 0.18 | 0.31 |
| $A_3$ | 0.16 | 0.06 | 0.20 | 0.03 |
| $A_4$ | 0.20 | 0.18 | 0.20 | 0.24 |
| $A_5$ | 0.22 | 0.29 | 0.26 | 0.17 |

Table 5.  Initial DMM X with values, which must be minimised, changed to maximised values

| Alternatives | Criteria | | | |
|---|---|---|---|---|
|  | $X_1$ | $X_2$ | $X_3$ | $X_4$ |
| Optimization Direction | MAX | MAX | MIN | MIN |
| Weight of criterion | **0.29** | **0.34** | **0.22** | **0.15** |
| $A_0$ | 1.00 | 1.00 | 6.67 | 33.33 |
| $A_1$ | 0.18 | 0.18 | 6.67 | 4.17 |
| $A_2$ | 0.24 | 0.29 | 5.56 | 3.23 |
| $A_3$ | 0.16 | 0.06 | 5 | 33.33 |
| $A_4$ | 0.20 | 0.18 | 5 | 4.17 |
| $A_5$ | 0.22 | 0.29 | 3.85 | 5.88 |

Table 6. Normalised DMM  $\overline{X}$

|  | $X_1$ | $X_2$ | $X_3$ | $X_4$ |
|---|---|---|---|---|
| W | **0.29** | **0.34** | **0.22** | **0.15** |
| $A_0$ | 0.50 | 0.50 | 0.02 | 0.40 |
| $A_1$ | 0.09 | 0.09 | 0.20 | 0.05 |
| $A_2$ | 0.12 | 0.15 | 0.17 | 0.04 |
| $A_3$ | 0.08 | 0.03 | 0.15 | 0.40 |
| $A_4$ | 0.10 | 0.09 | 0.15 | 0.05 |
| $A_5$ | 0.11 | 0.15 | 0.12 | 0.07 |

Table 7. Solution Result Weighted normalised DMM $X$ and final result.

|  | $X_1$ | $X_2$ | $X_3$ | $X_4$ | S | K | Rank |
|---|---|---|---|---|---|---|---|
| $A_0$ | 0.145 | 0.17 | 0.044 | 0.06 | 0.105 | 1 | |
| $A_1$ | 0.026 | 0.031 | 0.044 | 0.008 | 0.027 | 0.257 | 4 |
| $A_2$ | 0.034 | 0.051 | 0.037 | 0.006 | 0.032 | 0.304 | 1 |
| $A_3$ | 0.023 | 0.010 | 0.033 | 0.06 | 0.031 | 0.295 | 2 |
| $A_4$ | 0.029 | 0.031 | 0.033 | 0.008 | 0.025 | 0.238 | 5 |
| $A_5$ | 0.032 | 0.051 | 0.026 | 0.010 | 0.029 | 0.276 | 3 |

So $A_2 > A_3 > A_5 > A_1 > A_4$.

So among the five projects: $P_2 > P_3 > P_5 > P_1 > P_4$ and P2 is the best project among all five projects.

## 7. CONCLUSION

The traditional approaches of optimization used within the engineering context are based on assumption. The modelling of engineering problem is based on a different kind of logic, taking into consideration the existence of multicriteria, conflicting aims of decision maker, the complex nature of evaluation process.

 Above all, the main advantage of MCDM provides taking decision by analyzing complex problem; possibilities to aggregate criteria in evaluation process; chances of taking appropriate decisions; scope for decision maker to participate actively in the process of decision making.

According to the proposed method, the degree of alternative choice is made by comparison of variables which are analyzed with ideally best one.

In conclusion, the proposed method provides a simple approach of complex theory to access alternative projects and select the best set of project by using the described integrated approach of AHP and ARAS method. This integrated approach has a great future in project management field.

**Authors**

TULI BAKSHI

She works as an assistant professor in Computer Application department in Calcutta Institute of Technology, West Bengal, India. She possess three master degrees in Applied Mathematics, Computer Application and Information Technology

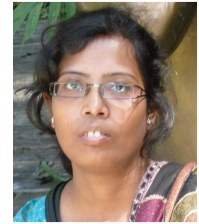

DR. BIJAN SARKAR

He is the Professor and Former Head of Production Engineering Department, Jadavpur University, Kolkata, India. He has received Outstanding Paper Award at Emerald Literati Network for Excellences 2006, UK. He had also received the Best paper Awards from "Indian Institute of Industrial Engineering (IIIE), Mumbai" for the year 2002 and 2003. He was also awarded Certificates of Merit by the "Institution of Engineers India" during 2001-02 and 2002-03. He is the Co-author of the book on Production Management published by AICTE,CEP, New Delhi. He has published more than 150 papers in the National / International Journals and Proceedings. He is the Reviewer of IJPR, EJOR, and IE (I). His fields of interests include Tribology, Reliability engineering, AI, Soft-computing applications and Decision engineering.

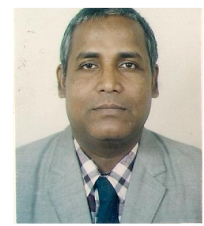